\documentclass[9pt,conference]{IEEEtran}
\usepackage{amssymb,amsthm,amsmath,array}
\usepackage{graphicx}
\usepackage[caption=false,font=footnotesize]{subfig}
\usepackage{xspace}
\usepackage[sort&compress, numbers]{natbib}
\usepackage{stmaryrd}
\usepackage{xcolor}
\usepackage{mathtools}
\usepackage{float}
\usepackage{textcomp}

\begin{document}
\title{Set projection algorithms for blind ptychographic phase retrieval}
\author{\IEEEauthorblockN{
        Wenjie Mei\IEEEauthorrefmark{1}, 
        Andrew M. Maiden\IEEEauthorrefmark{2}
    }
    \IEEEauthorblockA{
        \IEEEauthorrefmark{1}\IEEEauthorrefmark{2} Dept. of Electrical and Electronic Engineering, University of Sheffield, Sheffield, S1 3JD.\\
        \IEEEauthorrefmark{2} Diamond Light Source, Harwell Science and Innovation Campus, Fermi Ave, Didcot OX11 0DE.\\
        }
}
\maketitle
\begin{abstract}
Set projection algorithms are a class of algorithms used in ptychography to help improve the quality of the reconstructed images. In this paper, existing and new projection algorithms for ptychography are described and compared.
\end{abstract}

\section{Introduction}

Set projection algorithms play a critical role in the reconstruction process of ptychography, and their performance greatly affects the accuracy and quality of the final images. Different set projection algorithms combine and iterate between the projections onto constraint sets in different ways, generally, these projections can be defined by a relaxed projection notation $P^a_Sx$, where $S$ indicates the constraint set and $a$ indicates the relaxation. When $a=1$, $P^1_Sx$ represents the standard projection of $x$ onto $S$. In terms of the standard projection, relaxed projection can be written as $P^a_Sx=(aP^1_S+(1-a)I)x$, where $I$ is the identity operator. When $0<a<1$, it moves $x$ only a fraction of the distance toward $P^1_S$, which is under-relaxed projection, another way is over-relaxed projection, where $a>1$ and moves $x$ beyond $P^1_S$. There is a special case called reflection about a constraint set, where $a=2$, $R_Sx=P^2_Sx=(2P^1_S-I)x$, which moves $x$ twice as far as $P^1_S$ in the same direction. A geometric example of three non-convex constraint sets $S,T,U$ is illustrated in Fig. 1. The arrows from $x$ in Fig.1 show several different projections with different degrees of relaxation.

\begin{figure}[htbp]
    \centering
    \includegraphics[scale=0.7]{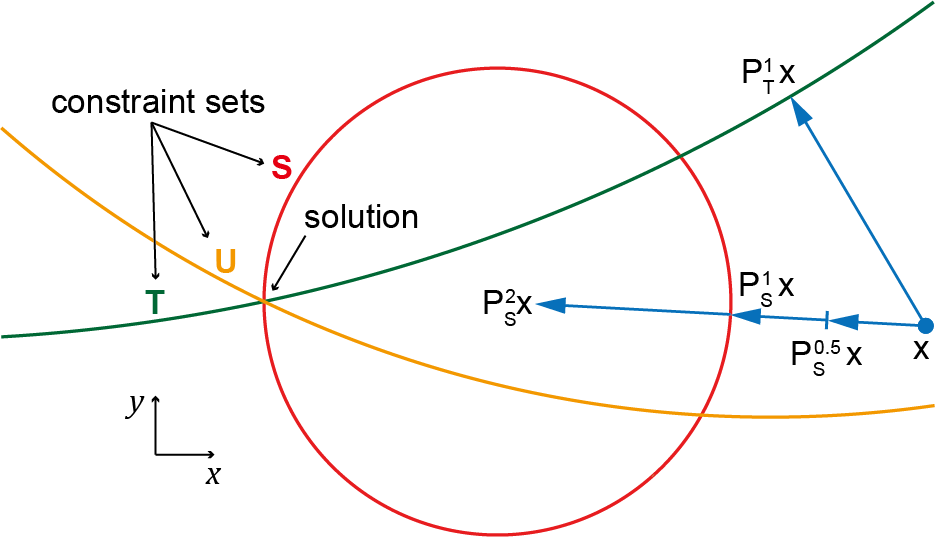}
    \caption{Projection schematic}
    \label{fig:my_label}
\end{figure}

To find the intersection of constraint sets $S,T,U$, a single projection is clearly not enough, the Sequential Projections (SP) algorithm is  the most intuitive way to combine these projection operations into an algorithm to solve this problem. As shown in Fig.2, given an initial start point $x_0$, the algorithm projects onto one of the constraints $S$, takes the result of this projection and projects it onto a second constraint $T$, finally, projects onto constraint $U$, the result of this sequence of projections is the point $x_1$ in Fig.2. Then repeats, the next iteration starts from point $x_1$, the order of projections can either be fixed or with the order shuffled at random between iterations, moreover, the relaxation of each projection also can be changed during sequential projections for better performance.

\begin{figure}[htbp]
    \centering
    \includegraphics[scale=0.7]{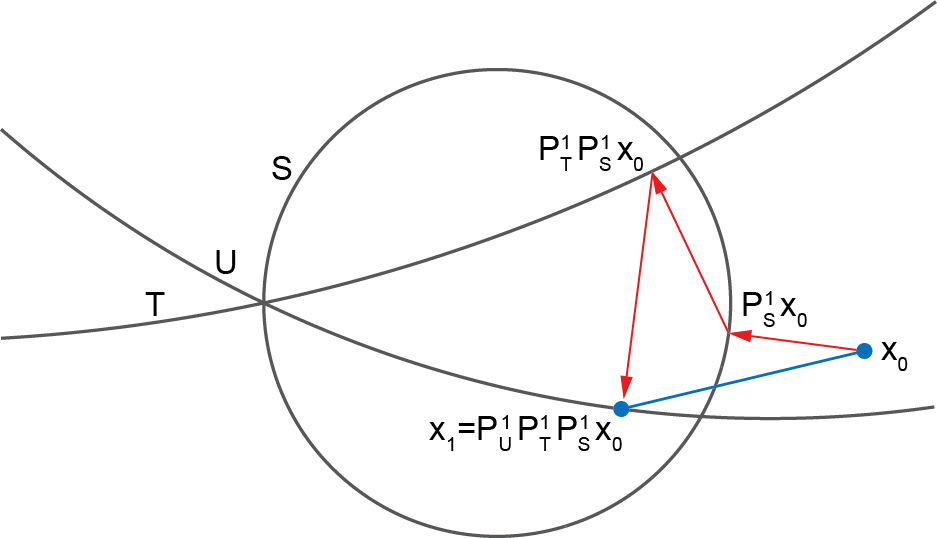}
    \caption{Sequential Projections (SP) schematic}
    \label{fig:my_label}
\end{figure}

However, because the Sequential Projections algorithm cycles through the constraint sets sequentially, it can become entrenched in periodic oscillations rather than reaching a settled point of convergence. This is especially true where the underlying data are noisy, so that the sets may not all intersect at a single solution point. Therefore, product space is used in many set projection algorithms to treat constraint sets collectively rather than sequentially to avoid this problem. The idea is that the project $x_k$ onto each of the constraint sets in parallel (the divide step), then these projections are averaged in some way to reach a single consensus solution (the concur step), the entire process is called "divide and concur"\cite{code2}. The schematic diagram is shown in Fig.3

\begin{figure}[htbp]
    \centering
    \includegraphics[scale=0.7]{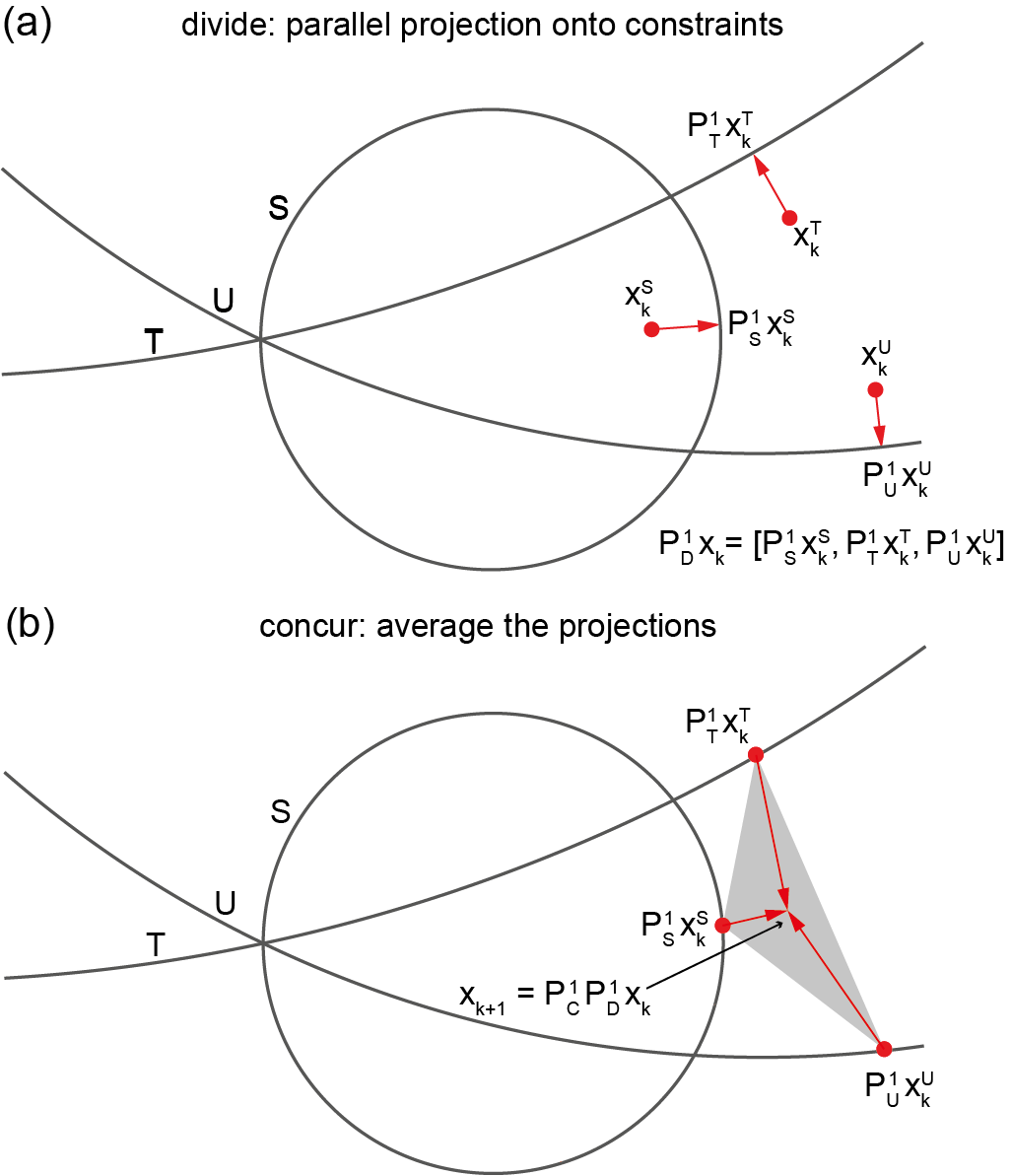}
    \caption{Schematic of "divide and concur"}
    \label{fig:my_label}
\end{figure}

The product space in this instance consists of three copies of the 2D plane, each holding one of the circle constraints. Each 2D plane also has its own estimate of the solution, so that $x_k$ has three components: $x_k=[x^S_k,x^T_k,x^U_k]$. The divide and concur approach begins with the divide step, shown in Fig.3(a), calculates the individual projection onto different sets. These parallel projections can be considered as a standard single 'divide' projection $P^1_D$ in the product space, giving three new points $P^1_Dx_k=[P^1_Sx^S_k,P^1_Tx^T_k,P^1_Ux^U_k]$. The next step is "concur", which averages the individual projections from the divide step. This averaging operation can also be expressed as a standard concur projection in the product space: $P^1_Cx_k=(x^S_k+x^T_k+x^U_k)/3$. Therefore, the divide and concur (DC) algorithm, alternates between the divide and concur projections, the next iteration will be $x_{k+1}=P^1_CP^1_Dx_k$. By changing the relaxation of each projection, the DC approach can also be generalised as the equation below. 
$$ x_{(k+1)}=aP^b_CP^c_Dx_k+(1-a)x_k  $$
Where $a$,$b$ and $c$ are relaxation parameters.

With appropriate choice of $a$, $b$ and $c$, this general formula can implement most major set projection algorithms. We investigate an overview of the main set projection algorithms used in ptychography, such as Divide and concur (DC), Average reflections (AR), Douglas Rachford (DR), Solvent flip (SF), Relaxed averaged alternating reflections (RAAR), Reflect reflect relax(RRR) and T$_\lambda$ algorithm\cite{code2}\cite{code1}\cite{code3}\cite{code4}\cite{code5}\cite{code6}\cite{code7}\cite{code8}\cite{code9}\cite{code10}.

Fig.4. is a general flow diagram of the ‘abc’ ptychographic method, the computation starts with an initial guess of probe and object, then forms exit waves for all of the scan positions, do the projections, and finally updates probe and object.

\begin{figure}[htbp]
    \centering
    \includegraphics[scale=0.43]{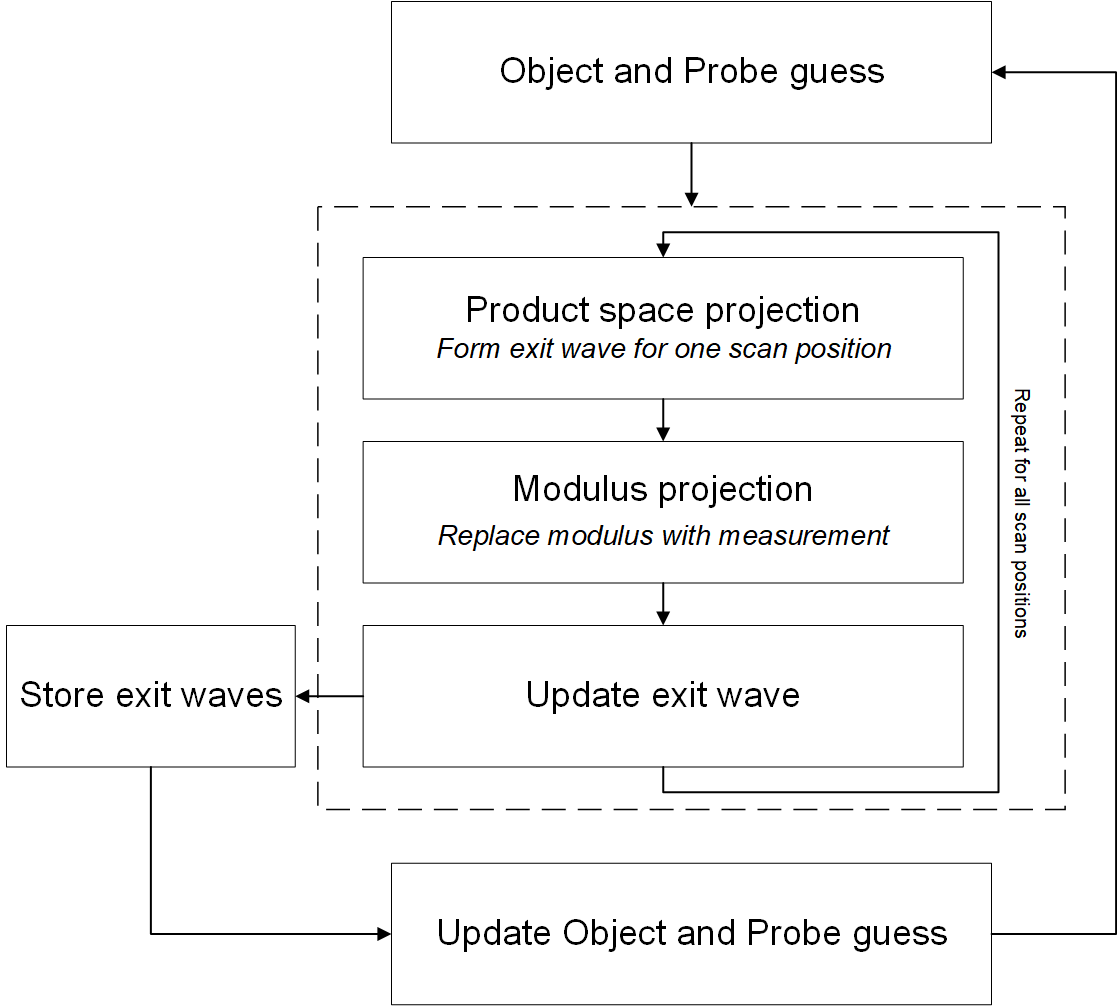}
    \caption{The flow chart of generalised set projection algorithm}
    \label{fig:my_label}
\end{figure}

A simulation on blood cells shows the different performance of these set projection algorithms, the object is shown in Fig. 5. and two different size of probes are shown in Fig.6.

\begin{figure}[h]
    \centering
    \includegraphics[scale=0.8]{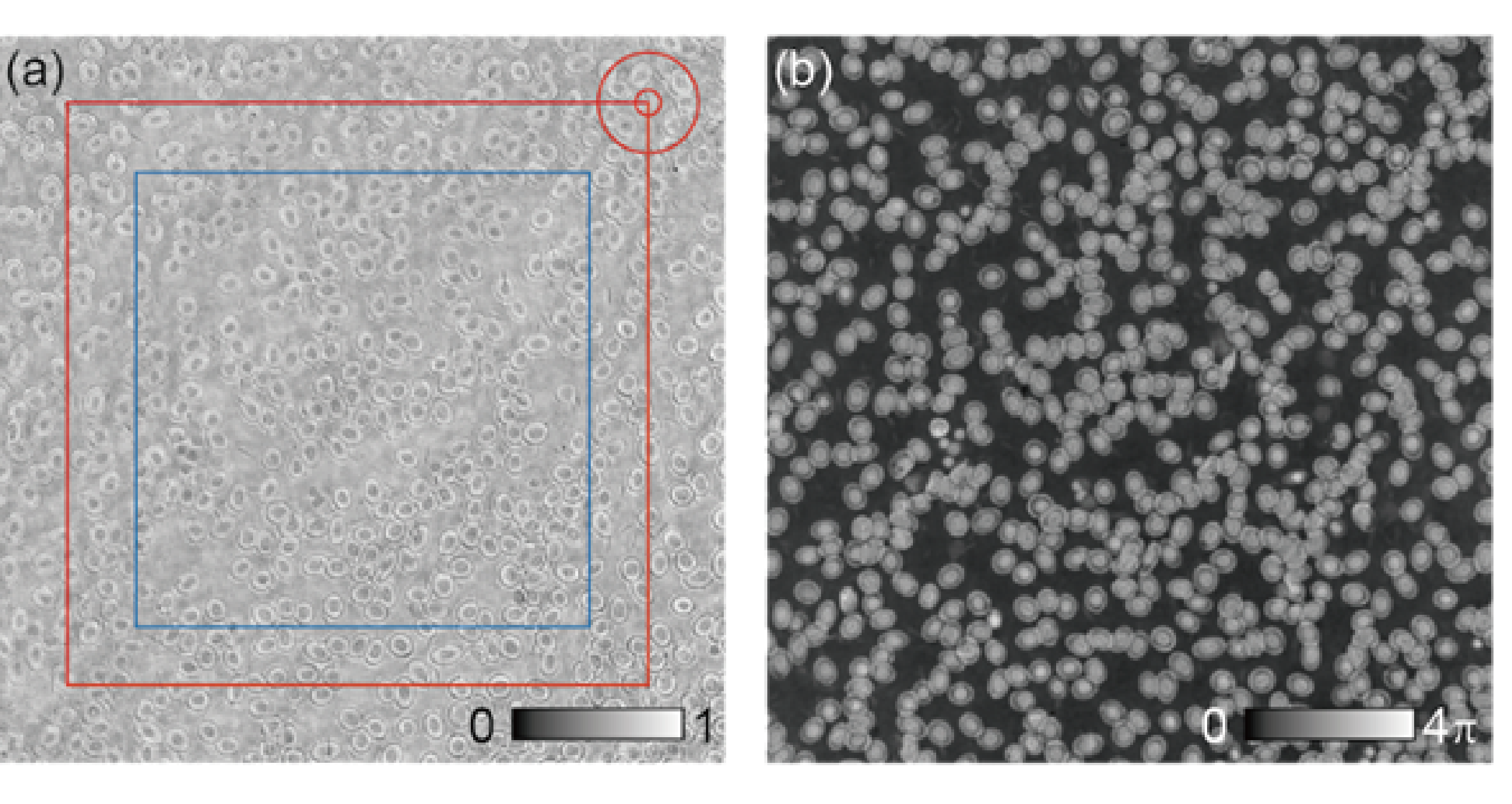}
    \caption{The simulation object, (a) the modulus of object, the red box indicates the reconstruction area, red circles indicates two different size of probe, and the blue box shows the area of error metric. (b) the phase of object.}
    \label{fig:my_label}
\end{figure}

\begin{figure}[h]
    \centering
    \includegraphics[scale=0.8]{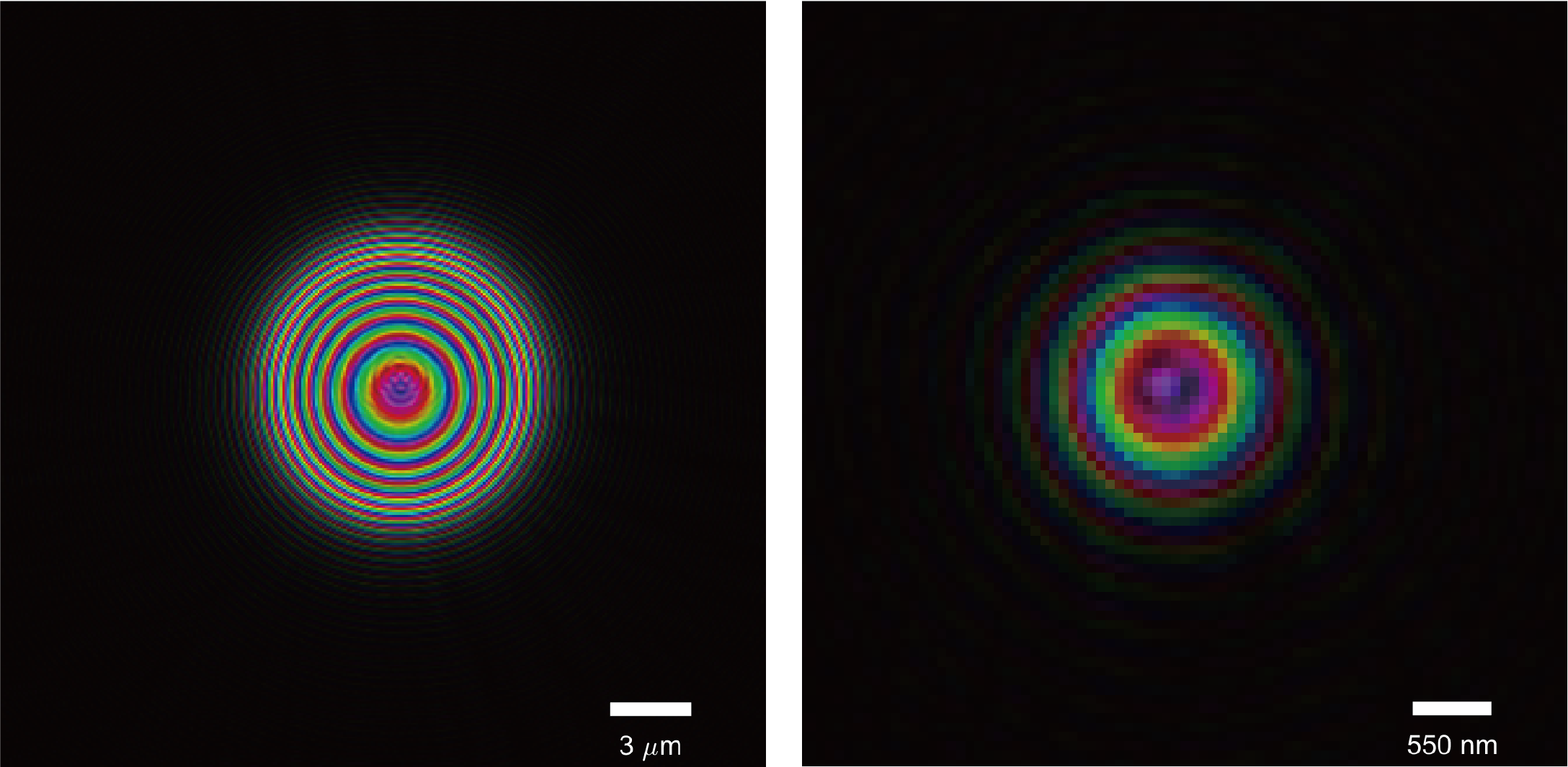}
    \caption{The simulation probes}
    \label{fig:my_label}
\end{figure}

The simulation results are shown in Fig.7. and Fig.8., which are the error metrics for different set projection algorithms.

\begin{figure}[h]
    \centering
    \includegraphics[scale=0.45]{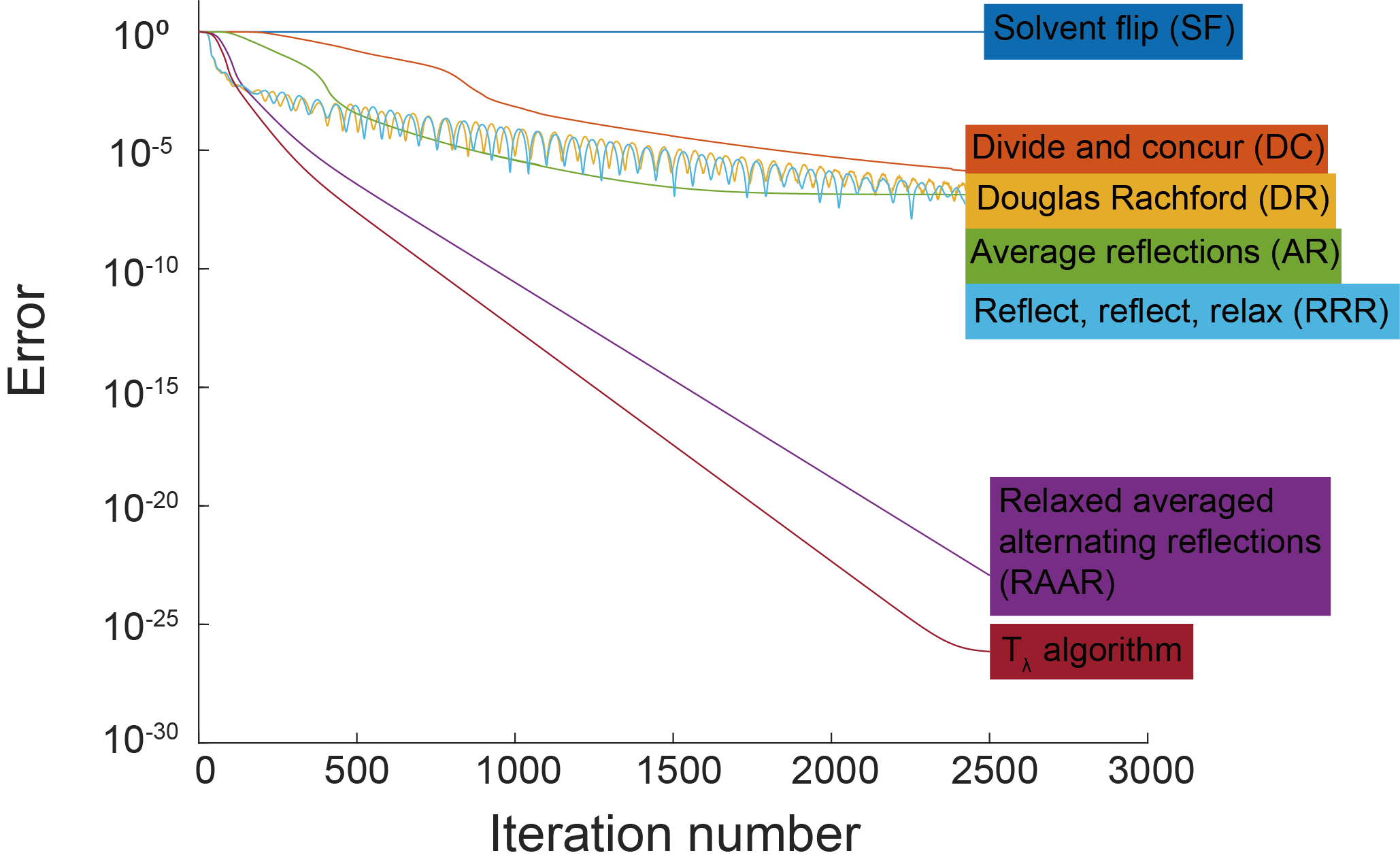}
    \caption{The error lines of simulation using big probe for set projection algorithms}
    \label{fig:my_label}
\end{figure}

\begin{figure}[h]
    \centering
    \includegraphics[scale=0.45]{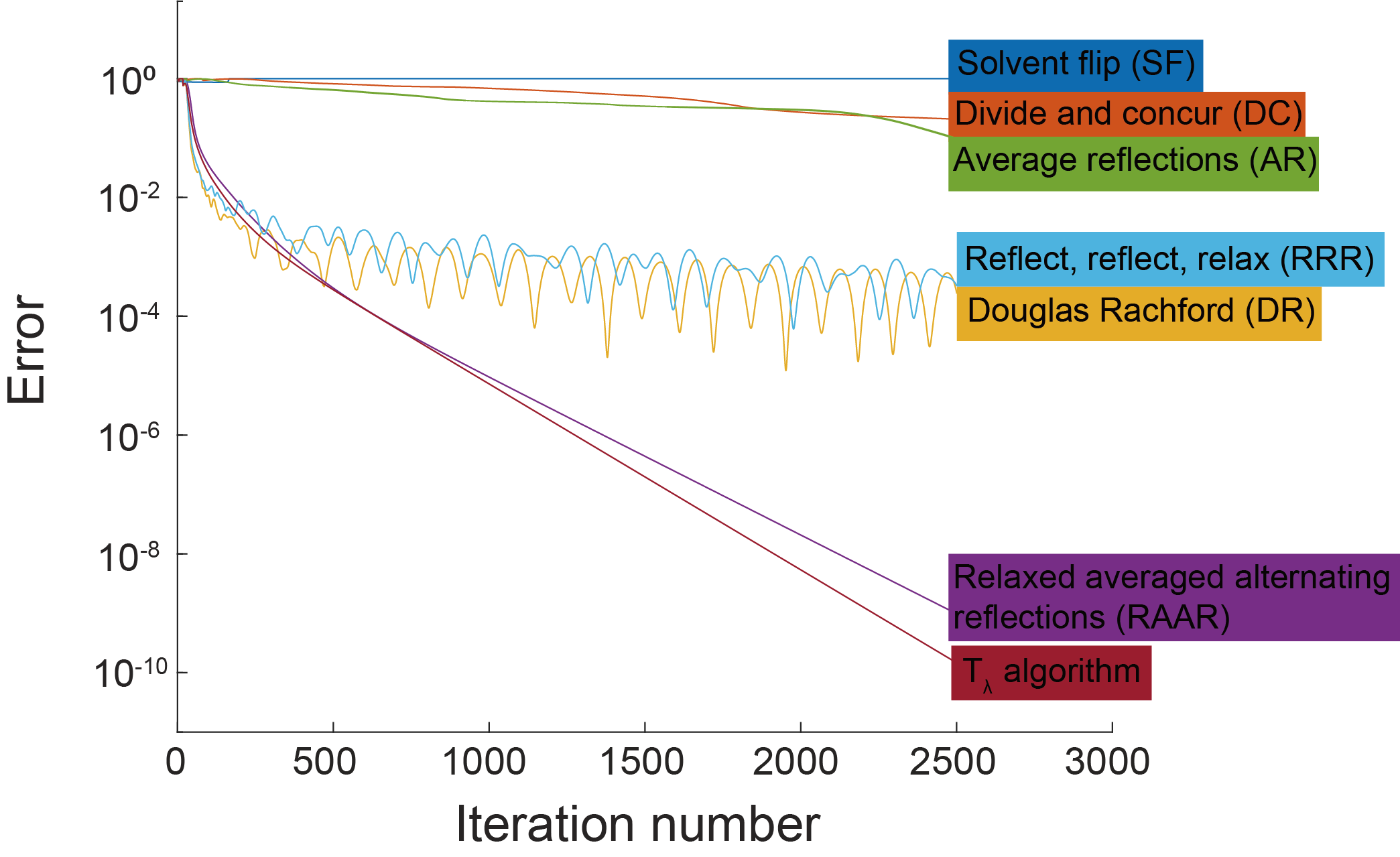}
    \caption{The error lines of simulation using small probe for set projection algorithms}
    \label{fig:my_label}
\end{figure}

\section{Conclusion}
In conclusion, based on the generalised 'abc' projection model,  several different ptychographic projection methods are implemented and compared by adjusting the values of relaxation parameters 'abc'. In our simulation tests, T$_\lambda$ algorithm is slightly better than Relaxed averaged alternating reflections (RAAR), and both of them have significantly better performance than other methods.

\clearpage
\newpage
\bibliographystyle{IEEEtran}
\bibliography{references}
\end{document}